\documentclass[12pt]{article}
\usepackage{geometry}
\newcommand{\beq}{\begin{equation}}
\newcommand{\eeq}{\end{equation}}
\newcommand{\beqs}{\begin{eqnarray}}
\newcommand{\eeqs}{\end{eqnarray}}

\begin{document}

\begin{titlepage}

\hfill ULB-TH/09-31

\vspace{40pt}

\begin{center}
{\Large \bf Why not a di-NUT? \\
\large or\\
Gravitational duality and rotating solutions }

\end{center}

\vspace{15pt}

\begin{center}
{\large  Riccardo Argurio and Fran\c cois Dehouck}\\

\vskip 35pt

Physique Th\'eorique et Math\'ematique and International Solvay Institutes \\
Universit\'e Libre de Bruxelles, C.P. 231, 1050 Bruxelles, Belgium
\end{center}

\vspace{20pt}

\begin{center}
\textbf{Abstract}
\end{center}
We study how gravitational duality acts on rotating solutions,
using the Kerr-NUT black hole as an example. After properly
reconsidering how to take into account both electric (i.e.
mass-like) and magnetic (i.e. NUT-like) sources in the equations
of general relativity, we propose a set of definitions for the
dual Lorentz charges. We then show that the Kerr-NUT solution has
non-trivial such charges. Further, we clarify in which respect
Kerr's source can be seen as a mass $M$ with a dipole of NUT
charges.
\end{titlepage}

%%%%%%%%%%%%%%%%%%%%%%%%%%%%%%%%%%%%%%%%%%

%%%%%%%%%%%%%%%%%%%%%%%%%%%%%%%%%%%%%%%%%%%%%%%%%%%%%%%%%%%%%%%%%%%%%%%%%%%%%%%%%%%%%%%%%%%%%
%%%%%%%%%%%%%%%%%%%%%%%%%%%%%%%%%%%%%%%%%%%%%%%%%%%%%%%%%%%%%%%%%%%%%%%%%%%%%%%%%%%%%%%%%%%%%
\section{Introduction}

The theory of general relativity, when linearized, shares many
similarities with the much simpler theory of electromagnetism. The
latter has the particular feature of having vacuum equations which are
invariant under a duality transformation, which exchanges electric and
magnetic fields. This invariance, extended to the case when sources
are present, implies the existence not only of electric charges, but
also of magnetic monopoles. The Dirac monopole \cite{Dirac:1931kp}, 
dual of the Coulomb
charge, is then interpreted as a source in the Bianchi identities of
the electromagnetic field strength rather than in its equations of motion.

Gravitational duality is the transposition to general relativity
of the same idea of duality \cite{Nieto:1999pn} (see
\cite{Henneaux:2004jw} for an Hamiltonian proof of the duality).
The vacuum equations are invariant under a duality which is
defined on the Riemann tensor. Extending this duality in presence
of sources, see \cite{Bunster:2006rt} and \cite{Barnich:2008ts},
implies the existence not only of matter giving rise to the
ordinary stress-energy tensor, but also of ``magnetic'' matter giving
rise to a dual stress-energy tensor. In particular, the
Schwarzschild solution must have, at least at the linearized
level, a dual solution, which was long ago identified with the
Lorentzian Taub-NUT solution. The literature on the subject is
vast, we list here a few references for the interested reader: for
generalizations of the duality to (A)dS space, see
\cite{Leigh:2007wf}--\cite{Julia:2005ze}; 
for duality of higher-spin field theories see
for example \cite{Hull:2001iu} and  \cite{Boulanger:2003vs} and
references therein; for considerations about extending the duality
to the full theory, see \cite{Julia:2005ze}--\cite{Bekaert:2002uh}.

One feature of the Taub-NUT solution is to have a string-like
singularity \cite{Misner:1963fr}, sometimes called the Misner string, much similar to the
Dirac string of the magnetic monopole. Then similarly, the Misner
string can be considered as a gauge artifact in the metric as soon as
one is ready to accept the presence of a ``magnetic'' source in the
r.h.s. of the cyclic identity for the Riemann tensor, or in other
words a magnetic stress-energy tensor.

When the string singularity is properly taken into account in this
way, it becomes possible in general relativity to define a surface
integral that computes precisely this ``magnetic'' NUT charge
\cite{Argurio:2008zt} (see also \cite{Bossard:2008sw} for an
approach using Komar charges).

It is the purpose of this note to extend these ideas to rotating
solutions. The simplest occurrence, which will be our main focus
below, is the solution obtained when performing a duality rotation
on the familiar Kerr black hole. The Kerr-NUT black hole \cite{DN} is a
subgroup of the general Petrov type D metrics obtained by
Plebanski and Demianski in \cite{Plebanski:1976gy} and it was shown to
be consistent with gravitational duality in
\cite{Turakulov:2001jc}. A global analysis of this solution can be
found in \cite{MILLER}.

It is then legitimate to ask what is the singularity
structure, or what are the sources, for such a solution, and what are
its charges and how to compute them by surface integrals at
infinity.\footnote{We will focus here and below only on solutions which
  are locally asymptotically flat.}

In the course of this investigation, we will see that there is
also a ``physical" choice involved. As the Dirac monopole could be
considered as a semi-infinite solenoid, one could provide a
similar physical interpretation of the Taub-NUT solution. This was
first reported by Bonnor in \cite{Bonnor} (see also
\cite{Manko:2005nm}). The choice becomes more tricky when dealing
with solutions like the linearized Kerr black hole and its dual
(see however \cite{MILLER} for some considerations along the lines
of \cite{Bonnor}). In fact, we show that one could also 
consider the source of the Kerr metric as made of an electric mass
$M$ and a pair of NUT sources, which we will refer as a di-NUT, in
some appropriate limit.

 The paper is organised as follows. In Section 2, we review the way
  Einstein equations were written in a duality invariant way in
  \cite{Bunster:2006rt} and we single out the fact that dualization
  looks more natural when realized on Lorentz indices (see also \cite{Bergshoeff:2008vc}). In Section 3,
  we derive the ADM and dual ADM charges when string
  contributions are included. We notice that there exists, in our formalism,
  no way of expressing the generalized Lorentz
  charges as surface integrals in a gauge-invariant way. In Section
  4, we study the particular case of the Kerr-NUT solution and
  show that the usual description of the Kerr metric as a rotating
  point source of mass $M$ could also be interpreted as a point source $M$
  with a monopole anti-monopole pair in the limit where the monopole
  mass goes to infinity and the distance in between them goes to zero
  while keeping the orbital momentum fixed. Appendix A recalls the
  duality between the linearized Schwarzschild and NUT solutions, along
  with their respective sources. Appendix B details the calculations
  we need for the interpretation of the sources of the Kerr-NUT
  metric.

%%%%%%%%%%%%%%%%%%%%%%%%%%%%%%%%%%%%%%%%%%%%%%%%%%%%%%%%%%%%%%%%%%%%%%%%%%%%%%%%%%%%%%%%%%%%%
%%%%%%%%%%%%%%%%%%%%%%%%%%%%%%%%%%%%%%%%%%%%%%%%%%%%%%%%%%%%%%%%%%%%%%%%%%%%%%%%%%%%%%%%%%%%%

\section{Gravitational duality on Lorentz indices}

In this section, we review how gravitational duality
works in linearized general relativity by re-deriving the duality invariant
form of the Einstein equations, cyclic and Bianchi identities.
We will argue that gravitational duality is best understood
when dualization is performed
on Lorentz indices. We show that this choice permits to lower the duality
relation to a duality between spin connections. We eventually give an expression of the spin connection in terms of the vielbein
and a three index object, first introduced in \cite{Bunster:2006rt},
that contains the magnetic information of the solution. 
Since we linearize around flat Minkowski
space in cartesian coordinates, there will be no distinction between
curved and flat indices in the following.

When there are no magnetic charges, the Einstein equations, cyclic and
Bianchi identities are:
\begin{eqnarray}
&& G_{ \mu \nu}= 8 \pi G T_{\mu \nu}, \nonumber \\
&& R_{\mu[\nu\alpha\beta]}= \frac{1}{3} ( R_{ \mu \nu \alpha
\beta} + R_{ \mu \beta \nu \alpha} + R_{ \mu \alpha \beta \nu})
= 0, \nonumber \\
&&\partial_{[\alpha}\: R_{| \rho \sigma | \beta \gamma]}=
\frac{1}{3} (
\partial_{\alpha}\: R_{\rho \sigma  \beta \gamma} +
\partial_{\gamma} \: R_{ \rho \sigma  \alpha  \beta} +
\partial_{\beta} \:  R_{ \rho \sigma \gamma \alpha} ) = 0.
\end{eqnarray}
The Bianchi identities are solved by expressing the Riemann tensor in
terms of a spin connection. In turn, the cyclic identity is solved
when the spin connection is expressed in terms of a vielbein or, when
the local Lorentz gauge freedom is fixed, in terms of a (linearized) metric.

Gravitational duality tells us that for every metric, there exists a
dual metric such that their respective Riemann tensors are dual to
each other. One important difference with electromagnetism and its
two-form field strength is that here the Riemann tensor has two pairs
of antisymmetric indices (the Lorentz and the form indices,
respectively, in reference to the spin connection) and a choice for
the duality relation should be made.
We will prefer here, for reasons to be explained later, a dualization
on the Lorentz indices (the first two indices in our conventions, as
is clear from the Bianchi identities above):
\begin{eqnarray}\label{gravdual}
\tilde{R}_{\mu\nu\rho\sigma}=\frac{1}{2} \:
\varepsilon_{\mu\nu\alpha\beta}R^{\alpha \beta}_{\:\:\:\:\:
\rho\sigma}, \:\:\:\:\:\:\:\: R_{\mu\nu\rho\sigma}=- \frac{1}{2}
\varepsilon_{\mu\nu\alpha\beta} \tilde{R}^{\alpha
\beta}_{\:\:\:\:\: \rho\sigma},
\end{eqnarray}
where $\tilde{R}_{\mu\nu\rho\sigma}$ denotes the magnetic or dual Riemann tensor.

Looking now at the magnetic cyclic identity, we have
\begin{eqnarray}
 ( \tilde{R}_{ \mu \nu \alpha \beta} + \tilde{R}_{ \mu
\beta \nu \alpha} + \tilde{R}_{ \mu \alpha \beta \nu}) &=& 3
\delta^{\rho\sigma \kappa}_{[\nu\alpha\beta]} \tilde{R}_{ \mu \rho
\sigma \kappa}
 =-\frac{1}{2} \varepsilon_{\gamma \nu
\alpha \beta} (\varepsilon^{\gamma \rho \sigma \kappa}\tilde{R}_{
\mu \rho \sigma \kappa} ) \nonumber \\
&=&-\frac{1}{2} \varepsilon_{\gamma \nu
\alpha \beta}( 2 R^{\gamma}_{\:\:\: \mu}-
\delta^{\gamma}_{\:\:\mu} R)= 8 \pi G \varepsilon_{ \nu \alpha \beta \gamma}  T^{\gamma}_{\:\:\mu}\label{rela}.
\end{eqnarray}
and we see that the duality makes the electric stress-energy tensor appear at the r.h.s. of the equation. However,
under a gravitational duality rotation
\begin{eqnarray}
&& R_{ \mu \nu \rho \sigma} \rightarrow  \tilde{R}_{ \mu \nu \rho
\sigma}, \qquad \tilde{R}_{ \mu \nu \rho \sigma} \rightarrow -R_{
\mu \nu \rho
\sigma},\nonumber \\
&& T_{\mu \nu} \rightarrow  \Theta_{\mu\nu}, \qquad \qquad \Theta_{\mu
\nu} \rightarrow -T_{\mu\nu},
\end{eqnarray}
meaning that the electric cyclic identity can be generalized such as to include
a magnetic stress-energy tensor $\Theta_{\mu\nu}$. We write the full set of electric and magnetic equations respectively as:
\begin{eqnarray}\label{emequations}
&& G_{ \mu \nu}= 8 \pi G T_{\mu \nu}, \nonumber \\
&& R_{ \mu \nu \alpha
\beta} + R_{ \mu \beta \nu \alpha} + R_{ \mu \alpha \beta \nu}
= - 8\pi G \varepsilon_{ \nu \alpha \beta
\gamma} \Theta^{\gamma}_{\:\:\mu}, \nonumber \\
&&\partial_{\epsilon}\: R_{\gamma \delta \alpha \beta} +
\partial_{\alpha} \: R_{ \gamma \delta  \beta \epsilon} +
\partial_{\beta} \:  R_{ \gamma \delta \epsilon \alpha} = 0, \nonumber \\
& & \nonumber \\
&& \tilde{G}_{ \mu \nu}= 8 \pi G \Theta_{\mu \nu}, \nonumber \\
&& \tilde{R}_{ \mu
\nu \alpha \beta} + \tilde{R}_{ \mu \beta \nu \alpha} +
\tilde{R}_{ \mu \alpha \beta \nu}
=  8\pi G \varepsilon_{ \nu \alpha \beta \gamma} T^{\gamma}_{\:\:\mu}, \nonumber \\
&&\partial_{\epsilon}\: \tilde R_{\gamma \delta \alpha \beta} +
\partial_{\alpha} \: \tilde R_{ \gamma \delta  \beta \epsilon} +
\partial_{\beta} \: \tilde  R_{ \gamma \delta \epsilon \alpha} = 0,
\end{eqnarray}
where the electric and magnetic cyclic identity can also be written by means
of (\ref{rela}) as
\begin{eqnarray}
\tilde{G}_{ \mu \nu}= 8 \pi G \Theta_{\mu \nu}, \qquad
G_{ \mu \nu}= 8 \pi G T_{\mu \nu},
\end{eqnarray}
thus showing the invariance of the equations under gravitational
duality rotation. One advantage of dualizing on Lorentz indices, as compared
to a dualization on form indices, is that we do not need to modify the Bianchi identity because:
\begin{eqnarray}
 \partial_{[\alpha}\: \tilde{R}_{| \mu \nu | \beta
\gamma]}=\frac{1}{2}
\varepsilon_{\mu\nu}^{\:\:\:\:\rho\sigma}\partial_{[\alpha}\: R_{|
\rho \sigma | \beta \gamma]}.
\end{eqnarray}
Note that the vanishing of the Bianchi identity is consistent with the
cyclic identity having a non-trivial source term if and only if the
magnetic stress-energy tensor is conserved, $\partial_\mu\Theta^{\mu
  \nu}=0$, just as the ordinary stress-energy tensor.

As already mentioned previously, the Riemann tensor can only be defined in terms of a metric
when both the cyclic and Bianchi identities have a trivial
right-hand side. To deal with the introduction of magnetic sources we
introduce, as in \cite{Bunster:2006rt}, a three-index object ${\Phi^{\mu\nu}}_{\rho}$ such that:
\begin{eqnarray}\label{PHI}
\partial_{\alpha} \Phi^{\alpha \beta}_{\:\:\:\:\:\gamma} = -16 \pi
G \Theta^{\beta}_{\:\:\:\gamma}, \qquad \Phi^{\alpha
\beta}_{\:\:\:\:\:\gamma}= - \Phi^{ \beta
\alpha}_{\:\:\:\:\:\gamma}.
\end{eqnarray}
Further we define
\begin{eqnarray}
\bar{\Phi}^{\rho\sigma}_{\:\:\:\:\:\alpha}=
\Phi^{\rho\sigma}_{\:\:\:\:\:\alpha} + \frac{1}{2} (
\delta^{\rho}_{ \:\:\: \alpha } \Phi^{\sigma} -\delta^{\sigma}_{
\:\:\: \alpha }\Phi^{\rho}), \qquad  \Phi^{\rho}= \Phi^{\rho
\alpha}_{\:\:\:\:\: \alpha}.
\end{eqnarray}
The Riemann tensor that will be solution of the set of equations (\ref{emequations}) when making use of (\ref{PHI})
 is:
\begin{eqnarray}\label{genriemann}
R_{\alpha \beta \lambda \mu} = r_{\alpha \beta \lambda \mu}+
\frac{1}{4} \: \epsilon_{\alpha \beta \rho \sigma}
(\partial_{\lambda} \bar{\Phi}^{\rho\sigma}_{\:\:\:\:\:\mu}
-\partial_{\mu} \bar{\Phi}^{\rho\sigma}_{\:\:\:\:\:\lambda} ),
\end{eqnarray}
where $r_{\alpha \beta \lambda \mu}$ is the usual Riemann tensor verifying the usual cyclic and
Bianchi identities with no magnetic stress-energy tensor. This means that $r_{\alpha \beta
\lambda \mu}= r_{\lambda \mu \alpha \beta } $ and that it can be
derived from a potential: $r_{\alpha \beta \lambda \mu}= 2
\partial_{[\alpha} h_{\beta][ \lambda,\mu]}$.

Another advantage of the dualization on Lorentz indices comes
directly from the vanishing r.h.s. of the Bianchi identity which
gives us the right to express the linearized Riemann tensor in
terms of a spin connection by
\begin{eqnarray}
R_{\mu\nu\rho\sigma}=\partial_{\rho} \omega_{\mu \nu \sigma}-
\partial_{\sigma} \omega_{\mu \nu \rho}, \label{spinco}
\end{eqnarray}
and thus allows to lower
the duality relation between Riemann tensors to a duality between
spin connections. With the help of (\ref{gravdual}) and (\ref{spinco}) the gravitational duality relation becomes:
\begin{eqnarray}\label{dualspinc}
\tilde{\omega}_{\mu \nu \sigma}= \frac{1}{2}
\varepsilon_{\mu\nu\alpha\beta} \: \omega^{\alpha \beta
}_{\:\:\:\:\:\sigma},
\end{eqnarray}
where this relation is true up to a gauge transformation as the spin connection is a gauge-variant object.

The linearized vielbein and spin connection for the Riemann
tensor $r_{\mu\nu\alpha\beta}$ are
\begin{eqnarray}
r_{\mu\nu\rho\sigma}&=& \partial_{\rho} \Omega_{\mu \nu \sigma}-
\partial_{\sigma} \Omega_{\mu \nu \rho}, \nonumber \\
e^\mu &=& dx^\mu+\frac{1}{2}
\eta^{\mu\nu}(h_{\nu\rho}+v_{\nu\rho})dx^\rho, \nonumber \\
\Omega_{\mu\nu} &=& \Omega_{\mu\nu\rho}e^\rho, \qquad
\Omega_{\mu\nu\rho}= \frac{1}{2}(\partial_\nu h_{\mu\rho}
-\partial_\mu h_{\nu\rho} +\partial_\rho v_{\nu\mu}),
\end{eqnarray}
where $h_{\mu\nu}=h_{\nu\mu}$ is the linearized metric and $v_{\mu\nu}=-v_{\nu\mu}$. Using this together with  relations (\ref{spinco}) and (\ref{dualspinc})  gives us the spin connection in terms of the vielbein and the three-index object $\Phi_{\mu\nu\rho}$:
\begin{eqnarray}\label{spinconnection}
\omega_{\mu\nu\rho}&=& \Omega_{\mu\nu\rho} +\frac{1}{4}
\varepsilon_{\mu\nu\gamma\delta} \bar{\Phi}^{\gamma
\delta}_{\:\:\:\:\: \rho} \nonumber \\
&=&\frac{1}{2}(\partial_\nu h_{\mu\rho} -\partial_\mu h_{\nu\rho}
+\partial_\rho v_{\nu\mu}) +\frac{1}{4}
\varepsilon_{\mu\nu\gamma\delta} \bar{\Phi}^{\gamma
\delta}_{\:\:\:\:\: \rho}.
\end{eqnarray}
In \cite{Argurio:2008zt} it was realized that by means of
(\ref{dualspinc}) there always exists a ``regular" spin
connection even when magnetic sources are present.\footnote{By
  ``regular" we should stress that we only refer to 
  (string) singularities on the two-sphere at spatial infinity.} From
the expression above this can be achieved for a 
specific choice of $v_{\mu\nu}$ that cancels string
contributions coming from $\Phi_{\mu\nu\rho}$.

One also easily sees that:
\begin{eqnarray}
\tilde{\omega}_{\mu\nu\sigma}&=&\frac{1}{2}
\varepsilon_{\mu\nu\alpha\beta} \: \omega^{\alpha \beta
}_{\:\:\:\:\:\sigma} = -\frac{1}{4} [ \varepsilon_{\mu\nu\alpha
\beta} (2 \partial^{\alpha}
h^{\beta}_{\:\:\:\sigma}+\partial_{\sigma} v^{\alpha\beta})+ 2
\bar{\Phi}_{\mu \nu \sigma}].
\end{eqnarray}

%%%%%%%%%%%%%%%%%%%%%%%%%%%%%%%%%%%%%%%%%%%%%%%%%%%%%%%%%%%%%%%%%%%%%%%%%%%%%%%%%%%%%%%%%%%%%
%%%%%%%%%%%%%%%%%%%%%%%%%%%%%%%%%%%%%%%%%%%%%%%%%%%%%%%%%%%%%%%%%%%%%%%%%%%%%%%%%%%%%%%%%%%%%

\section{The dual Poincar\'{e} charges}

In this section, we give the generalized expressions for the ADM
momenta and dual ADM momenta in presence of NUT charge. These were
first established in \cite{Argurio:2008zt} for a specific gauge
choice of the vielbein. Here, we give a full treatment of the singular
string contributions, obtaining gauge-independent expressions for the surface integrals. This is also a proof of the validity of the gauge choice of
\cite{Argurio:2008zt}.  We eventually apply the same idea to derive general
expressions for the Lorentz charges and their duals though we will show
that there is no way in this formalism to express
the charges as surface integrals without partially fixing the gauge.

The generalized ADM momenta and dual ADM momenta are:
\begin{eqnarray}
 P_\mu =\int T_{0\mu} d^3x \nonumber = \frac{1}{8\pi G} \int G_{0\mu}d^3x,  \\
K_\mu= \int\Theta_{0\mu} d^3x = \frac{1}{8\pi G} \int \tilde G_{0\mu}d^3x.
\end{eqnarray}
However, the electric and magnetic Einstein tensors can be expressed as \cite{Argurio:2008zt}:
\begin{eqnarray}\label{GGtilde}
G_{0\mu} &=&  \partial_i ({\omega^{0i}}_\mu
+\delta^0_\mu {\omega^{i\rho}}_\rho
-\delta^i_\mu {\omega^{0\rho}}_\rho ),  \nonumber \\
\tilde G_{0\mu} &=& \varepsilon^{ijk}\partial_i
\omega_{\mu jk},
\end{eqnarray}
where by convention $\varepsilon^{ijk}=-\varepsilon^{0ijk}$.
This enables us to formulate the momenta as surface integrals:
\begin{eqnarray}
P_\mu & = & \frac{1}{8\pi G} \oint [ {\omega^{0l}}_\mu
+\delta^0_\mu {\omega^{l\rho}}_\rho
-\delta^l_\mu {\omega^{0\rho}}_\rho ] d \Sigma_l, \\
K_\mu & = & \frac{1}{8\pi G} \oint  \varepsilon^{ljk}
\omega_{\mu jk}
 d \Sigma_l .
\end{eqnarray}
With the help of (\ref{spinconnection}), we have:
\begin{eqnarray}
P_0 & = & \frac{1}{16\pi G} \oint \biggl [ \partial_i h^{li}-  \partial^l {h^{i}}_{i}+ \partial_i v^{il}+\varepsilon^{ljk} \Phi_{0jk} \biggl ] d \Sigma_l, \label{p0v}\\
P_k & = & \frac{1}{16\pi G} \oint \biggl [ \partial_0 {h^{l}}_{k}- \partial^l
h_{0k} + \delta^l_k \partial^{i} h_{0i} - \delta^l_k \partial_{0}
{h^{i}}_{i}
+\partial_{k} {v_{0}}^{l}+ \delta^l_k \partial^{i} v_{i 0} \nonumber \\
&& \:\:\:\:\:\:\:\:\: \:\:\:  - \frac{1}{2}\varepsilon^{lij}[ \Phi_{ijk}+\delta_{ik} {\Phi_{j0}}^{0}+\delta_{ik }{\Phi_{jm}}^{m}]+ \frac{1}{2}\delta^{l}_{k} \varepsilon^{ijm}\Phi_{ijm} \biggl ] d \Sigma_l, \label{pkv}\\
K_0 & = & \frac{1}{16\pi G} \oint \biggl [ \varepsilon^{lij}  [ \partial_i h_{0j}+ \partial_j v_{i0} ]+{\Phi^{l0}}_{0} \biggl ]  d \Sigma_l , \\
K_k & = & \frac{1}{16\pi G} \oint  \biggl [ \varepsilon^{lij}  [
\partial_i h_{kj}+ \partial_j v_{ik} ]+{\Phi^{l0}}_{k} \biggl ]  d
\Sigma_l.
\end{eqnarray}

When there are no magnetic charges, $\Theta_{\mu\nu}$ is zero and
thus $K_{\mu}$ also by definition. Setting ourselves in the gauge
where $v_{\mu\nu}=0$ one easily recognizes the ADM momenta
$P_{\mu}$. The important difference with electromagnetism is that
here the surface integrals for calculating the charges depend on
the spin connection, a gauge-variant object. In
electromagnetism the contribution of the Dirac string is
always equal to the opposite of the string contribution coming from the
regularized connection. Here, if we want to cancel the string contributions
we need the additional gauge freedom of the
vielbein to be fixed in the right gauge. By
duality arguments we showed that such a choice is always possible.
This completes the proof of the validity of the expressions
used in \cite{Argurio:2008zt}.
Details of calculations can be found in Appendix A.

In the same spirit, the general
expression for the Lorentz charges and their duals are as
follows:\footnote{Note that the fixed timelike index is now upstairs,
  contrary to the definitions of the momenta. We hope that this
  (arbitrary but innocuous) switch
  in the convention will not upset the reader too much.}
\begin{eqnarray}
L^{\mu\nu}&=& \int   (x^{\mu} T^{0 \nu }- x^{\nu} T^{0 \mu
}) d^3 x =\frac{1}{8\pi G} \int  (x^{\mu} G^{0 \nu }- x^{\nu} G^{0
\mu }) d^3 x,
\nonumber \\
\tilde{L}^{\mu\nu}&=& \int (x^{\mu} \Theta^{0 \nu }- x^{\nu}
\Theta^{0 \mu }) d^3 x = \frac{1}{8\pi G} \int (x^{\mu}
\tilde{G}^{0 \nu }- x^{\nu} \tilde{G}^{0 \mu}) d^3 x.
\end{eqnarray}

Plugging the expression (\ref{GGtilde}) into the definition of the electric Lorentz charges
leads us to:
\begin{eqnarray}\label{volumecharges}
L^{ij}&=& \frac{1}{8\pi G}\int (x^{i} G^{0 j}- x^{j} G^{0 i}) d^3 x
\nonumber \\
      &=& \frac{1}{8\pi G} \oint  \biggl [ x^{j} [\omega^{0li}-
      \delta^{li}{\omega^{0 k}}_{k}]- x^{i} [\omega^{0lj}-
      \delta^{lj}{\omega^{0 k}}_{k}] \biggl ] d\Sigma_l  +
      \frac{1}{8\pi G}\int  [\omega^{0ij}- \omega^{0ji}] \: d^3 x, \nonumber \\
L^{0i}&=&\frac{1}{8\pi G}\int (t G^{0 i}- x^{i} G^{0 0}) d^3 x
\nonumber
\\
      &=& \frac{1}{8\pi G} \oint \biggl [ -t [ \omega^{0li}-\delta^{li} {\omega^{0 k}}_{k} ]- x^{i} {\omega^{lj}}_{j}  \biggr] d\Sigma_l  +
      \frac{1}{8\pi G}\int  {\omega^{ij}}_{j} \: d^3 x.
\end{eqnarray}
We see that in the presence of non-trivial $\Phi_{\mu\nu\rho}$, we
have a priori no way to express the charges as surface integrals.
However, we know that the charges are independent of the choice of
$v_{\mu\nu}$, one could then always try to choose a gauge such as
to cancel the $\Phi_{\mu\nu\rho}$ contributions present in the
volume integrals by choosing an appropriate $v_{\mu\nu}$ . Expanding the volume integrals in the above expressions:
\begin{eqnarray}
\int 2 {\omega^{ij}}_{j} \: d^3 x &=& \int [ \partial_j
h^{ij}-\partial^{i} {h^{j}}_{j}+\partial_{j}v^{ji}+
\varepsilon^{ijk} \Phi_{0jk} ] \: d^3 x, \nonumber \\
\int 2 [\omega^{0ij}- \omega^{0ji}] \: d^3 x &=&  \int [
\partial^{i} h^{0j}  -  \partial^{j} h^{0i} +   \partial^{j} v^{i0} -
\partial^{i} v^{j0} - \varepsilon^{ijk} \Phi_{k0}^{\:\:\:\:\: 0} ]
\: d^3 x,
\end{eqnarray}
where we simplified the last equation using the relation
 $ \varepsilon^{lk[i} \bar{\Phi}_{lk}^{\:\:\:\:\:
j]}= \varepsilon^{ijk} \Phi_{k0}^{\:\:\:\:\: 0}$,
we see that we can absorb the $\Phi_{\mu\nu\rho}$ by choosing the $v_{ij}$ and the $v_{0i}$ such that:
\begin{eqnarray}
\int \partial_{j}v^{ij} \: d^3 x &=& \int \varepsilon^{ijk}
\Phi_{0jk} \: d^3 x,\label{gauge1} \\
\int [\partial^{j}v^{i0} - \partial^{i}v^{j0}] \: d^3
x &=&  \int \varepsilon^{ijk} \Phi_{k0}^{\:\:\:\:\: 0} \:
d^3 x.\label{gauge2}
\end{eqnarray}
Actually, these gauge choices do not fix completely the local
Lorentz gauge, and hence $v_{\mu\nu}$. Rather, they restrict the
gauge to a choice satisfying the above integral relations. Of
course this can be done in the simplest way by choosing a
$v_{\mu\nu}$ that locally compensates the singularity contained in
$\Phi_{\mu\nu\rho}$.

In the gauge choice of expressions (\ref{gauge1}) and
(\ref{gauge2}), we now have:
\begin{eqnarray}
L^{ij} &=& \frac{1}{8\pi G} \oint \biggl [ x^{j} [\omega^{0li}-
      \delta^{li}{\omega^{0 k}}_{k}]- x^{i} [\omega^{0lj}-
      \delta^{lj} {\omega^{0 k}}_{k}] +\frac{1}{2} [\delta^{il} h^{0j}-\delta^{jl} h^{0i}] \biggl ] d\Sigma_l, \nonumber
      \\
L^{0i} &=& \frac{1}{8\pi G} \oint \biggl [ -t [
\omega^{0li}-\delta^{li} {\omega^{0 k}}_{k} ]- x^{i}
{\omega^{lj}}_{j}  +  \frac{1}{2} [h^{il}-\delta^{il} h]\biggr]
d\Sigma_l.
\end{eqnarray}

If we now look at the dual Lorentz charges, we
have:
\begin{eqnarray}\label{volumedualcharges}
\tilde{L}^{0i} &=& \frac{1}{8\pi G}\int (t \tilde {G}^{0 i}- x^{i}
\tilde{G}^{0 0}) d^3 x \nonumber
\\ &=& \frac{1}{8\pi G} \oint  -\varepsilon^{ljk}
[ t \:  {\omega^{i}}_{jk} + x^{i} \omega_{0jk}]   d\Sigma_l  +
\frac{1}{16\pi G} \int \varepsilon^{ikl}
[\omega_{0kl}-\omega_{0lk}]d^3 x
 \nonumber \\
\tilde{L}^{ij}&=& \frac{1}{8\pi G}\int (x^{i} \tilde {G}^{0 j}-
x^{j} \tilde {G}^{0 i}) d^3 x
\nonumber \\
&=&\frac{1}{8\pi G}  \oint \varepsilon^{lkm} [ x^{j}
{\omega^{i}}_{km} - x^{i} {\omega^{j}}_{km} ] d\Sigma_l +
\frac{1}{8\pi G} \int \varepsilon^{ijk} {{\omega_{k}}^{l}}_{l} d^3
x
\end{eqnarray}
where in the last equality we used $\varepsilon^{ikm}
{\omega^{j}}_{km}-
\varepsilon^{jkm}{\omega^{i}}_{km}=\varepsilon^{ijk}
{{\omega_{k}}^{l}}_{l}$.

It is amusing to observe that the pieces in $L_{\mu\nu}$ and $\tilde
L_{\mu\nu}$ that cannot be expressed as surface integrals actually enjoy
a duality relation, $\tilde L_{\mu\nu}^\mathrm{bulk}=\frac{1}{2}
\varepsilon_{\mu\nu\rho\sigma}L^{\rho\sigma}_\mathrm{bulk}$. This
surprising property cannot of course be extended to the full charges,
as is obvious from their definition in terms of the stress-energy
tensor and its dual, respectively. However, a consequence of this
observation is that with the previous choice of gauge, we can also
express the dual charges as surface integrals:
\begin{eqnarray}
\tilde{L}^{0i}&=& \frac{1}{8\pi G} \oint \biggr [ -\varepsilon^{ljk}
[ t \: {\omega^{i}}_{jk} + x^{i} \omega_{0jk}]+\frac{1}{2}
\varepsilon^{ilk} h_{0k} \biggl] d\Sigma_l \nonumber \\
\tilde{L}^{ij}&=& \frac{1}{8\pi G}  \oint \biggr [\varepsilon^{lkm} [
x^{j} {\omega^{i}}_{km} - x^{i} {\omega^{j}}_{km} ] + \frac{1}{2}
\varepsilon^{ijk}[{h^{l}}_{k}-\delta^{l}_{k}h] \biggl] d\Sigma_l.
\end{eqnarray}

The expressions derived here for the electric and magnetic Lorentz charges are thus valid
in whatever gauge when expressed as volume integrals like in (\ref{volumecharges}) and (\ref{volumedualcharges}). Moreover, we have shown that there
exists a gauge choice valid for the Lorentz charges and their dual
that permits to eliminate the
$\Phi_{\mu\nu\rho}$ and simplify the expressions to surface
integrals. Note that in the case that all the $\Phi_{\mu\nu\rho}$ would be zero, whatever gauge
is obviously fine, and we recover the ADM expressions.

We are now prepared to apply those formulas to the Kerr-NUT solution.
Actually, it will
prove more efficient to work out the sources of the
solution, encoded in $T_{\mu\nu}$ and $\Theta_{\mu\nu}$, and compute
the charges from their original definition. The above arguments ensure
that the surface integrals, with a correct choice of gauge, will yield
the same result.

%%%%%%%%%%%%%%%%%%%%%%%%%%%%%%%%%%%%%%%%%%%%%%%%%%%%%%%%%%%%%%%%%%%%%%%%%%%%%%%%%%%%%%%%%%%%%
%%%%%%%%%%%%%%%%%%%%%%%%%%%%%%%%%%%%%%%%%%%%%%%%%%%%%%%%%%%%%%%%%%%%%%%%%%%%%%%%%%%%%%%%%%%%%

\section{Kerr and di-NUT sources}

There exists in the literature a generalization of the Taub-NUT
metric with three parameters, the ADM mass $M$, the NUT charge
$N$, and a rotation parameter $a$. This solution is known as the
Kerr-NUT metric. It is a particular case of the general
Petrov type D solution found in \cite{Plebanski:1976gy}. It was shown in
\cite{Turakulov:2001jc} that this metric is consistent with
gravitational duality. What we want to
study here are the different possible sources for the
linearized metric. We will see that to obtain a magnetic
stress-energy tensor such as the one for Kerr, we will need to
introduce the Misner string contribution in $\Phi_{\mu\nu\rho}$ which
appears in the surface integrals for $P_\mu$ and $K_\mu$ but also
point-like (Dirac delta) contributions.\\

The Kerr-NUT metric reads:
\begin{eqnarray}\label{MNa}
ds^2&=&-\frac{\lambda^2}{R^2} [dt- (a \sin^2\theta -2 N \cos\theta) d\phi]^2 +
\frac{\sin^2\theta}{R^2} [(r^2+a^2+N^2) d\phi - a dt]^2 \nonumber \\
&& +\frac{R^2}{\lambda^2} dr^2 +R^2 d\theta^2,
\end{eqnarray}
where $\lambda^2 = r^2-2Mr+a^2-N^2$ and  $R^2 = r^2+(N+a
\cos\theta)^2$.
We now consider some specific cases.\\

\noindent
{\it Taub-NUT ($a=0$)}

If we set $a=0$ in the above solution, we recover the Taub-NUT solution:
\begin{eqnarray}\label{MN}
ds^2&=&-\frac{\lambda^2}{R^2} [dt + 2 N \cos\theta d\phi]^2 +\frac{R^2}{\lambda^2} dr^2 +R^2 (d\theta^2+\sin^2\theta  d\phi^2),
\end{eqnarray}
where $\lambda^2 = r^2-2Mr-N^2$ and  $R^2 = r^2+N^2$. We review, in
Appendix A, the well-known duality that brings the linearized
Schwarzschild ($N=0$) to the linearized NUT solution ($M=0$). We also
see that the linearized NUT metric is actually to be supplemented with
the term ${\Phi^{0z}}_{0}=-16\pi N \delta(x)\delta(y)\vartheta(z)$ to
describe a source that is a point of magnetic mass
$N$.\footnote{Actually, in order for the string to be along the
  positive $z$ axis, we need to implement the change of coordinates
  $t\to t+2N \phi$ in the above metrics. This will always be
  assumed when referring to singularities. We refrain from
  implementing it on the explicit metrics to avoid unnecessary complications.}
If we do not
add this string contribution, the singularity is physical (as considered in \cite{Bonnor}) and can be
interpreted as a semi-infinite source of angular momentum
$\Delta L^{xy}=N\Delta z$.\\

\noindent
{\it Kerr ($N=0$)}

If we set $N=0$ in the metric (\ref{MNa}), we recover the Kerr
metric in Boyer-Lindquist coordinates:
\begin{eqnarray}\label{Ma}
ds^2=-(1-\frac{2Mr}{\Sigma}) dt^2 -\frac{4M a r }{\Sigma}
\sin^2\theta dt d\phi +\frac{\Sigma}{\Delta} dr^2 +\Sigma d\theta^2
+ \frac{ B}{\Sigma} \sin^2\theta d\phi^2,
\end{eqnarray}
where $ \Delta  \equiv  \lambda^2(N=0) = r^2-2Mr+a^2$, $\Sigma
\equiv  R^2 (N=0) = r^2+ a^2 \cos^2\theta$, and $B = (r^2+a^2)^2-
\Delta a^2 \sin^2\theta$. The charges of this metric are easily
calculated. If we linearize this metric at first order in the
charges, meaning we only keep terms in $M$ and $Ma$, we obtain:
\begin{eqnarray}\label{flucKerr1}
h_{00}=\frac{2M}{r}, \qquad  h_{ij}=\frac{2M}{r^3}x_{i}x_{j}, \qquad
h_{0i}= \frac{2Ma}{r^3} \varepsilon_{zij} x^{j}.
\end{eqnarray}
It is then shown in Appendix B.1. that, starting from the information about the metric, the source for this solution
is a rotating mass $M$ with angular momentum $J_z=L^{xy}=Ma$.\\

\noindent
{\it Rotating NUT ($M=0$)}

A more interesting metric is the one where we set $M$ to zero  in (\ref{MNa}). This is the rotating NUT metric.
Again, linearizing as before gives us:
\begin{eqnarray}
\tilde{h}_{tx}=\frac{2Nyz}{r (x^2+y^2)}, \:\:\:\:\:\:
\tilde{h}_{ty}=\frac{-2Nxz}{r (x^2+y^2)}, \:\:\:\:\:\:
\tilde{h}_{\mu\mu}=\frac{2Naz}{r^3}.
\end{eqnarray}
It is shown in Appendix B.2. that this linearized metric (after we set the string along the positive $z$-axis) supplemented with the $\Phi_{\mu\nu\rho}$ contributions:
\begin{eqnarray}
{\Phi^{0z}}_{0}&=&-16\pi N \delta(x) \delta(y) \vartheta(z),
\nonumber
\\
{\Phi^{0y}}_{x}&=& -{\Phi^{0x}}_{y}=-
{\Phi^{xy}}_{0}={\Phi^{yx}}_{0}= 8\pi N a \delta(\textbf{x}),
\end{eqnarray}
where $\vartheta$ is the usual Heaviside function, describes the
dual solution to the linearized Kerr. This means it describes a
point of magnetic mass $N$ and a magnetic angular momentum
$\tilde{L}^{xy}=Na$.\\

Let us recall now
that the choice for the $\Phi_{\mu\nu\rho}$ in the case of the
Taub-NUT solution found its meaning in the existence of a string singularity in the
linearized metric. 
This is also justified by considering the Schwarzschild metric as
electric and imposing gravitational duality.
Here, for the Kerr-NUT solution, one should note
that some $\Phi_{\mu\nu\rho}$ terms are only singular in $r=0$. 
Besides duality, we do not have any a priori argument in favour of
adding these delta contributions to the rotating NUT solution. One
could think of the linearized rotating NUT with only the string
contribution ${\Phi^{0z}}_{0}$ as another physical solution.
As shown in Appendix B.3, this would imply the presence of singular
terms in the electric stress-energy tensor corresponding to a dipole
of a positive and a negative mass at infinitesimal distance.
This interpretation is to be rejected on physical grounds because of
the presence of a negative mass in the compound.

It is on the other hand amusing to contemplate the dual situation, i.e.
the usual Kerr solution, where however we insert a non-trivial
magnetic stress-energy  tensor so that the non-trivial charges become
$P_0=M$ and $\tilde{L}_{0z}=Ma$. The
sources for this solution are:
\begin{equation}
T_{00}=M\delta(\textbf{x}), \qquad \Theta_{00}=Ma \delta(x)\delta(y)\delta'(z),
\end{equation}
 an electric point of mass $M$ and a di-NUT,
a dipole of NUT charges $+N$ and $-N$, separated by a distance $\epsilon$
when we take the limit $\epsilon \to 0$ and  $N \to \infty$ but with the
product $N\epsilon$  constant and equal to $\tilde L_{0z}=N\epsilon=Ma$:
\begin{eqnarray}
\Theta_{00}
&=& lim_{\epsilon \rightarrow 0}  [ N \delta(x)\delta(y)\delta(z+\epsilon/2)- N\delta(x)\delta(y)\delta(z-\epsilon/2)] \nonumber \\
&=& Ma  \delta(x)\delta(y)\delta'(z).
\end{eqnarray}
This situation is physical since there is no obstruction in having
negative NUT charges. Indeed, the Taub-NUT metrics with opposite signs
of $N$ are just related by a flip of the sign of the $\phi$ variable.
We should however note that this leads seemingly to a clash between the
statement of gravitational duality and positivity of the mass for the
Schwarzschild solution. In other words, according to 
the above arguments the gravitational dual of a
physical situation is not necessarily physical. It would be nice to
understand better this issue, with the use for instance of positive
energy theorems.

Concerning the euclidean Kerr black hole,
this interpretation had already been noticed a long time ago in
\cite{Gibbons:1979xm}. For the Lorentzian signature, it has
recently been observed in \cite{Manko:2009xx} that the Kerr metric
could be reproduced by a non-linear superposition of two Taub-NUT
black holes of opposite NUT charges.\footnote{We would like to
thank A. Virmani and R. Emparan for pointing out this reference to
us. } 
Here, we have clarified that if this is indeed true from the
perspective of the metrics, there is nevertheless a
difference depending on whether the $\delta'$ singularities find themselves
in the $T_{0i}$ components of the ordinary stress-energy tensor or in 
the $\Theta_{00}$ component of the magnetic dual. The
difference is encoded in the tensor $\Phi_{\mu\nu\rho}$
and is reflected on which Lorentz charges are non-trivial,
the electric or the magnetic ones. We suggest to identify the Kerr metric 
as a di-NUT only in the case where there is a non-trivial $\Theta_{00}$.

\subsection*{Acknowledgments}
We would like to thank G.~ Barnich, A.~ Kleinschmidt, L.~Houart,
C.~ Troessaert, A.~Virmani and V.~Wens for interesting
discussions. This work was supported in part by IISN-Belgium
(conventions 4.4511.06, 4.4505.86 and 4.4514.08) and by the
Belgian Federal Science Policy Office through the Interuniversity
Attraction Pole P5/27. R.A. is a Research Associate of the Fonds
de la Recherche Scientifique--F.N.R.S. (Belgium).

%%%%%%%%%%%%%%%%%%%%%%%%%%%%%%%%%%%%%%%%%%%%%%%%%%%%%%%%%%%%%%%%%%%%%%%%%%%%%%%%%%%%%%%%%%%%%
%%%%%%%%%%%%%%%%%%%%%%%%%%%%%%%%%%%%%%%%%%%%%%%%%%%%%%%%%%%%%%%%%%%%%%%%%%%%%%%%%%%%%%%%%%%%%
\appendix
\section{Taub-NUT}

In this appendix, we review the duality between the linearized
Schwarzschild and the linearized NUT solution with the conventions
set in Section 2. We recover both the ideas of Misner
\cite{Misner:1963fr} and Bonnor \cite{Bonnor} as how to interpret
the Taub-NUT solution. To deal with the Taub-NUT metric, Misner
noticed in \cite{Misner:1963fr} the presence of a string
singularity. Considering it as non-physical, he identifies time to
get rid of it. We show in Appendix A.2 that by gravitational
duality the string singularity in fact determines a
magnetic stress-energy tensor and is thus non-physical in an
``electric" theory. We do not discuss the identification as this
is really some feature that should be treated in the full theory.
If we drop this contribution, the magnetic stress-energy
tensor is zero and we end up with a massless source of angular
momentum $N$ at every point along the physical singularity at
$\theta=0$. This is Bonnor's interpretation of the Taub-NUT
solution. The string is considered as a physical singularity in
the ``electric" theory. This is presented in section A.3.

\subsection{The linearized Schwarzschild solution}

Considering the Schwarzschild solution, the non-trivial components
of the linearized metric and spin connection are:
\begin{eqnarray}
h_{tt}&=&\frac{2M}{r}, \qquad h_{ij}=\frac{2M}{r^3}x_{i}x_{j},
\nonumber \\
\omega_{0i0}&=&\frac{1}{2} \partial_i h_{00}= -M \frac{x_i}{r^3}, \nonumber \\
\omega_{ijk}&=&\frac{1}{2} (\partial_j h_{ik}- \partial_i
h_{jk})=\frac{M}{r^3}(\delta_{jk}x_i-\delta_{ik}x_j).
\end{eqnarray}
The non-trivial components of the linearized Riemann tensor are:
\begin{eqnarray}
R_{0i0j}&=& - \partial_j \omega_{0i0}=  M( - \frac{3 x_i x_j}{r^5}
+ \frac{\delta_{ij}} {r^3}+ \frac{4\pi }{3} \delta_{ij}
\delta(\textbf{x})),
\nonumber \\
R_{ijkl}&=&  \partial_k \omega_{ijl} - \partial_l \omega_{ijk}
\nonumber \\ &=&  (\frac{2M}{r^3}+\frac{8\pi M}{3}
\delta(\textbf{x})) (\delta_{ik} \delta_{jl}- \delta_{il}
\delta_{jk} )\nonumber \\ && - \frac{3M}{r^5} (\delta_{ik} \:
x_{j} \:  x_{l}- \delta_{jk} \: x_{i} \:  x_{l} -\delta_{il} \:
x_{j} \: x_{k} + \delta_{jl} \: x_{i} \:  x_{k}),
\end{eqnarray}
where we used:
\begin{eqnarray}
\partial_j \frac{x_k}{r^3}&=& \frac{\delta_{jk}}{r^3}-\frac{3x_k
x_j}{r^5}+\frac{4\pi}{3} \delta_{jk} \delta(\textbf{x}).
\end{eqnarray}
We finally obtain: $R_{00}= 4\pi M \delta(\textbf{x})$, $ R_{ij} =
4\pi M \delta_{ij} \delta(\textbf{x})$ and $R =  8\pi M
\delta(\textbf{x})$. This is also $ G_{00}= 8 \pi T_{00}= 8 \pi M
\delta(\textbf{x}) $, $G_{ij}= T_{ij}=0$ and $G_{0j}= T_{0j}=0$.
The source for linearized Schwarzschild is thus a point of mass
$M$.

\subsection{The NUT solution from the dual Schwarzschild}
To obtain the ``electric" NUT spin connection, we use the duality
relation $\omega_{\mu\nu\sigma}=-\frac{1}{2}
\varepsilon_{\mu\nu\alpha\beta}\:
{\tilde{\omega}^{\alpha\beta}}_{\:\:\:\:\:\sigma}$ where
$\tilde{\omega}$ is the spin connection for the linearized
Schwarzschild after we applied the duality rotation
$\omega\rightarrow \tilde{\omega}$ and $M\rightarrow N$. We thus
%\begin{eqnarray}
%\tilde{\omega}_{0i0}= -N \frac{x_i}{r^3} \:\:\:\:\:\:
%\tilde{\omega}_{ijk}=\frac{N}{r^3}(\delta_{jk}x_i-\delta_{ik}x_j),
%\end{eqnarray}
obtain the regular spin connection for the NUT solution:
\begin{eqnarray}
\omega_{ij0}= \varepsilon_{ijk} \tilde{\omega}_{0k0}= -N
\varepsilon_{ijk} \frac{x_k}{r^3}, \qquad
\omega_{0ij}=-\frac{1}{2} \varepsilon_{ikl} \tilde{\omega}_{klj}=
N \varepsilon_{ijk} \frac{x_k}{r^3}.
\end{eqnarray}
This gives the non-trivial components of the Riemann tensor:
\begin{eqnarray}
R_{0i0j}&= & 0, \qquad R_{ijkl}=0, \nonumber \\
R_{ij0k} &=& N \varepsilon_{ijl} \partial_k (\frac{x^l}{r^3}) = N
\varepsilon_{ijl} (\frac{\delta_{kl}}{r^3}-\frac{3x_k x_l}{r^5}
+\frac{4\pi}{3} \delta_{kl} \delta(\textbf{x})), \nonumber \\
R_{0ijk} &=& \partial_j \omega_{0ik}- \partial_k \omega_{0ij}
\nonumber \\
         &=& -2N \varepsilon_{ijk} (\frac{1}{r^3} +
         \frac{4\pi}{3} \delta(\textbf{x})) +
         3N (\varepsilon_{ijl} \frac{x_k x_l}{r^5}-
         \varepsilon_{ikl} \frac{x_j x_l}{r^5}).
\end{eqnarray}
For the Einstein equation, we have trivially $R_{00}=R_{ij}=0$.
From the expressions above, one easily sees that
$R_{0i}=R_{i0}=0$. This means that $T_{\mu\nu}=0$. Plugging the
expressions in the cyclic identity, we obtain:
\begin{eqnarray}\label{cyclicNUT}
R_{0ijk}+R_{0kij}+R_{0jki}&=&-8 \pi \varepsilon_{ijk} \Theta^{00} \nonumber \\
  &=& -2N \varepsilon_{ijk} (\frac{3}{r^3} +
         4\pi \delta(\textbf{x}))    + 6N (\varepsilon_{ijl} \frac{x_k x_l}{r^5}-
         \varepsilon_{ikl} \frac{x_j x_l}{r^5}-
         \varepsilon_{kjl} \frac{x_i x_l}{r^5}), \nonumber \\
R_{00ij}+R_{0j0i}+R_{0ij0}&=& -8\pi \varepsilon_{ijk}
{\Theta^{k}}_{0},
\nonumber \\
 R_{i0jk}+R_{ik0j}+R_{ijk0}&=&-\partial_j
 (\omega_{0ik}+\omega_{ik0})+\partial_k
 (\omega_{0ij}+\omega_{ij0})
= -8\pi \varepsilon_{jkl}{\Theta^{l}}_{i}. \nonumber \\
\end{eqnarray}
This gives us:
\begin{eqnarray}
\Theta^{00}=N \delta(\textbf{x}), \qquad \Theta^{0k}=0, \qquad
\Theta^{li}=0.
\end{eqnarray}
For a solution describing a magnetic particle of mass $N$, and
thus a magnetic stress-energy tensor $\Theta^{00}=
N\delta(\textbf{x})$, we need, using relation (\ref{PHI}):
\begin{eqnarray}
{\Phi^{0z}}_{0}=-16\pi N \delta(x) \delta(y) \vartheta(z).
\end{eqnarray}
The previous non-trivial spin connections are expressed as:
\begin{eqnarray}
\omega_{ij0}=\frac{1}{2}(\partial_j h_{0i}-\partial_i h_{0j})+\frac{1}{4} \varepsilon_{ij0k}\: {\Phi^{0k}}_{0}, \nonumber\\
\omega_{0ij}=\frac{1}{2}(\partial_i h_{0j}+\partial_j
v_{i0})-\frac{1}{4} \varepsilon_{0ijk} \:  {\Phi^{0k}}_{0},
\end{eqnarray}
where we only assumed that the linearized vielbein is independent
on time. As we have established that the regular spin connection
is such that $\omega_{ij0}=-\omega_{0ij}$, we immediately see that
the right gauge fixing will be $h_{0i}=-v_{i0}$. The previous spin
connections are recovered with:
\begin{eqnarray}
h_{0x}=v_{0x}= 2N \frac{y}{r(r-z)}, \qquad v_{0y}=h_{0y}= - 2N
\frac{x}{r(r-z)},
\end{eqnarray}
where the metric has a singularity on the positive $z$-axis, in
agreement with the form of the $\Phi_{z00}$ term. To check that
this is the right result, we use a standard regularization
procedure (also used in the context of the Dirac monopole, see
e.g. \cite{Felsager:1981iy}):
\begin{eqnarray}
\vec{A}&=& (h_{0x}, h_{0y}, h_{0z}),\nonumber \\
\vec{B}&=& \vec{\nabla} \times \vec{A}= 2N
\frac{\vec{r}}{r^3}-8\pi N\delta(x)\delta(y)\vartheta(z) \hat{z},
\end{eqnarray}
where $\hat{z}$ is the unit vector along the $z$-axis and then:
\begin{eqnarray}
\partial_j h_{0i}-\partial_i h_{0j}= -
2 N \varepsilon_{ijk} \frac{x^{k}}{r^3}+ \varepsilon_{zij} 8\pi
N\delta(x)\delta(y)\vartheta(z).
\end{eqnarray}

Eventually note that the non-trivial contribution to the
linearized metric in spherical coordinates is:
\begin{eqnarray}
h_{0\phi}= - 2N(1+\cos\theta),
\end{eqnarray}
which is also the only non-trivial component for the linearized
NUT metric.

%With the vielbein supposed to be independent on time, the non-trivial spin connections are:
%\begin{eqnarray}
%\omega_{ij0}=\frac{1}{2}(\partial_j h_{0i}-\partial_i h_{0j})+\frac{1}{4} \varepsilon_{ij0k}({\Phi^{0k}}_{0}+{\Phi^{kl}}_{l}) \nonumber\\
%\omega_{0ij}=\frac{1}{2}(\partial_i h_{0j}+\partial_j v_{i0})+\frac{1}{4} \varepsilon_{0ikl} \bar{\Phi}^{kl}_{\:\:\:\: j}
%\end{eqnarray}

As previously said, this partially meets up with Misner's
interpretation of the Taub-NUT metric. Here, we interpret the
singularity at $\theta=0$ as non-physical in an ``electric" way
but it contributes to the magnetic stress-energy tensor. The
solution describes a particle of magnetic mass $N$.

\subsection{The NUT solution without the string}
To recover Bonnor's interpretation, we set to zero the
${\Phi^{\mu\nu}}_{\rho}$. Then, we obviously have
$\Theta_{\mu\nu}=0$. With the previous choice of $v_{\mu\nu}$, the
non-trivial components of the spin connections are now:
\begin{eqnarray}
\omega_{ij0}= -N \varepsilon_{ijk} \frac{x_k}{r^3}+
\varepsilon_{zij} 4\pi N\delta(x)\delta(y)\vartheta(z),
 \nonumber \\
\omega_{0ij}= N \varepsilon_{ijk}
\frac{x_k}{r^3}-\varepsilon_{zij} 4\pi
N\delta(x)\delta(y)\vartheta(z).
\end{eqnarray}
Note that we still have $\omega_{ij0}=-\omega_{0ij}$ so that from
(\ref{cyclicNUT}) we still immediately see that
$\Theta^{il}=\Theta^{0i}=0$. We can check that $\Theta^{00}=0$ as
it should be.

The non-trivial components for the Einstein tensor are:
\begin{eqnarray}
G_{i0}=-\partial_j (\varepsilon_{zij} 4\pi N
\delta(x)\delta(y)\vartheta(z)),
\end{eqnarray}
giving the non-trivial components of $T_{\mu\nu}$:
\begin{eqnarray}
T_{x0} =  -\frac{N}{2} \delta(x)\delta'(y)\vartheta(z), \qquad
T_{y0} = \frac{N}{2}  \delta'(x)\delta(y)\vartheta(z).
\end{eqnarray}
Note that such  $T_{\mu\nu}$ is conserved.

This shows that $P_\mu=0$  and $\Delta L^{xy}/\Delta z=N$ for every value along
the singularity. This agrees with Bonnor's interpretation of
the NUT solution as a massless source of angular momentum at the
singularity $\theta=0$.

\section{Kerr-NUT metric}

We now want to generalize the analysis of appendix A to the case
of the Kerr-NUT solution presented in section 4. We will see here
that the dual Kerr solution possesses the usual Misner string but
also additional delta contributions to the $\Phi_{\mu\nu\rho}$. If
we include these contributions, we get by gravitational duality a
magnetic mass $N$ with a magnetic angular momentum $J_z=Na$. If we
do not, we see that it corresponds to a dipole of electric masses
$M$ separated by a distance $\epsilon$ in the limit where
$M\rightarrow \infty$, $\epsilon \rightarrow 0$ but
$L_{0z}=M\epsilon=Na$ is constant. We only present the additional
information not contained in the previous Taub-NUT example as the
non-trivial contributions of the Kerr-NUT metric split into
contributions that were already present in the Taub-NUT case and
additional contributions in $Ma$ or $Na$.
%Obviously we
%could also implement the previous considerations here. If we add
%the delta contributions without adding the Misner string, the
%Kerr-NUT solution would then be the superposition of the Kerr
%geometry and a massless source of angular momentum at $\theta=0$

\subsection{Kerr metric}
The additional non-trivial components of the linearized metric and
linearized spin connection are:
\begin{eqnarray}\label{flucKerr2}
h_{0i}&=& \frac{2Ma}{r^3} \varepsilon_{zij} x^{j}, \nonumber \\
\omega_{0ij}&=& \frac{1}{2} \partial_i h_{0j}= -Ma
\varepsilon_{zij}(\frac{1}{r^3}+\frac{4\pi}{3}
\delta(\textbf{x}))- \frac{3Ma}{r^5} \varepsilon_{zjl}x_i x^l,
\nonumber \\
\omega_{ij0}&=& \frac{1}{2} (\partial_j h_{i0}-\partial_{i}
h_{j0})= \omega_{0ji}- \omega_{0ij} \nonumber \\
&=& Ma \varepsilon_{zij}(\frac{2}{r^3}+\frac{8\pi}{3}
\delta(\textbf{x}))- \frac{3Ma x^l}{r^5}
(\varepsilon_{zil}x_j-\varepsilon_{zjl}x_i).
\end{eqnarray}
The additional non-trivial components of the linearized Riemann
tensor are:
\begin{eqnarray}
R_{0ijk}&=&-Ma  \varepsilon_{zkl}(\partial_j \partial_i \partial_l
\frac{1}{r}) + Ma  \varepsilon_{zjl}(\partial_k \partial_i
\partial_l \frac{1}{r}), \nonumber \\
R_{ij0k}&=& -Ma  \varepsilon_{zjl}(\partial_k \partial_i
\partial_l \frac{1}{r}) + Ma  \varepsilon_{zil}(\partial_k
\partial_j
\partial_l \frac{1}{r}), \nonumber \\
\end{eqnarray}
where one can show that:
\begin{eqnarray}
\partial_i \partial_j \partial_k \frac{1}{r}& =& -15 \frac{x_i x_j
x_k}{r^7}+ \frac{3}{r^5} (\delta_{ij} x_k +\delta_{ki} x_j +
\delta_{jk} x_i)\nonumber \\
&&- \frac{4 \pi}{5} (\delta_{ij} \partial_k \delta(\textbf{r})+
\delta_{ki} \partial_j \delta(\textbf{r}) + \delta_{jk} \partial_i
\delta(\textbf{r})).
\end{eqnarray}
Combining these results with the ones from Appendix A, we easily
obtain: $R_{j0}=R_{0j}= {R_{0ij}}^{i}=Ma
\varepsilon_{zjl}(\partial_l \Delta \frac{1}{r})=-4\pi M a
\varepsilon_{zjl}\partial_l \delta(\textbf{x})$. This also gives
us: $R_{00}= 4\pi M \delta(\textbf{x})$, $ R_{ij} = 4\pi M
\delta_{ij} \delta(\textbf{x})$, $R =  4\pi M \delta(\textbf{x})$.
Eventually, we find: $ G_{00}= 8 \pi T_{00}= 8 \pi M
\delta(\textbf{x}) $, $G_{ij}= T_{ij}=0$ and $G_{0j}= R_{0j}=8 \pi
T_{0j}=-4\pi M a \varepsilon_{zjl}\partial_l \delta(\textbf{x}) $.
This solution describes a point of electric mass $M$ with an
electric angular momentum $L^{xy}=Ma$.

\subsection{The rotating NUT solution from the dual Kerr}
As for the dual of linearized Schwarzschild, by duality rotation
we obtain the additional spin connections of the dual Kerr metric:
\begin{eqnarray}
\omega_{0i0}&=& -\frac{1}{2} \varepsilon_{ijk}
\tilde{\omega}_{jk0}= Na \frac{\delta_{zi}}{r^3} -\frac{8}{3}
\pi N a \delta_{zi} \delta(\textbf{x}) - 3 N a \frac{z x_i}{r^5}, \nonumber \\
\omega_{ijk} &=& \varepsilon_{ijl} \tilde{\omega}_{0lk} = Na (
\delta_{zi} \delta_{kj} -\delta_{zj}\delta_{ki})(-\frac{2}{r^3}+
\frac{4\pi}{3} \delta(\textbf{x})) \nonumber \\
&&\:\:\:\:\:\:\:\:\:\:\:\:\:\: +\frac{3Na }{r^5}(x_k(x_j
\delta_{zi}-x_i \delta_{zj})+ z (x_i \delta_{kj}-x_j
\delta_{ki})),
\end{eqnarray}
where we used $\varepsilon_{ijk} \varepsilon^{zjk} = 2
\delta_{i}^{z}$  and $\varepsilon_{ijk} \varepsilon^{zjl} =
\delta_{z}^{i} \delta_{l}^{k} - \delta_{z}^{k}\delta_{l}^{i}$. One
can easily derive the Einstein tensor and find that this solution
corresponds to a magnetic point of mass $N$ with a magnetic
angular momentum $\tilde{L}^{xy}=Na$. This is the gravitational
dual of the Kerr solution with a $\Theta_{\mu\nu}$ with a
structure equal to the stress-energy tensor for Kerr, meaning:
\begin{eqnarray}
\Theta^{00}=N\delta(\textbf{x}), \qquad \Theta^{0x}= \frac{Na}{2}
\partial_{y} \delta(\textbf{x}), \qquad \Theta^{0y}= -
\frac{Na}{2} \partial_{x} \delta(\textbf{x}).
\end{eqnarray}
The non-trivial components for $\Phi_{\mu\nu\rho}$ are:
\begin{eqnarray}
{\Phi^{0z}}_{0}&=&-16\pi N \delta(x) \delta(y) \vartheta(z)
\nonumber
\\
{\Phi^{0y}}_{x}&=& -{\Phi^{0x}}_{y}=-
{\Phi^{xy}}_{0}={\Phi^{yx}}_{0}= 8\pi N a \delta(\textbf{x}).
\end{eqnarray}
We have:
\begin{eqnarray}
\omega_{0i0}&=&\frac{1}{2} \partial_i h_{00} + \frac{1}{4}
\varepsilon_{0ijk} {\Phi^{jk}}_{0}= \frac{1}{2} \partial_i h_{00}+
\frac{1}{2} \delta_{iz} {\Phi^{xy}}_{0}, \nonumber \\
\omega_{ijk}&=& \frac{1}{2} (\partial_j h_{ik}- \partial_i h_{jk}+
\partial_k v_{ji}) + \frac{1}{2} \varepsilon_{ij0l}
\bar{\Phi}^{0l}_{\:\:\:\: k},
\end{eqnarray}
where for our choice of $\Phi_{\mu\nu\rho}$:
\begin{eqnarray}
\frac{1}{2} \varepsilon_{ij0l} \bar{\Phi}^{0l}_{\:\:\:\: k}=
\frac{1}{2} \varepsilon_{ij0l}
{\Phi^{0l}}_{k}=(\delta_{iz}\delta_{jk}-\delta_{jz}\delta_{ik}){\Phi^{0y}}_x.
\end{eqnarray}
We then easily obtain:\footnote{Note that the $v_{\mu\nu}$ obtained
  here, and which lead to a regular spin connection, do not satisfy
  the gauge fixing proposed in Section 3, where the aim was rather to
  define surface integrals.}
\begin{eqnarray}
h_{00}= \frac{2Naz}{r^3}, \qquad h_{ij}= \frac{2Naz}{r^3}
\delta_{ij}, \qquad v_{ij}= \frac{2Na}{r^3}(\delta_{zi}
x_j-\delta_{zj}x_i).
\end{eqnarray}
The non-trivial components of the linearized metric in spherical
coordinates are then:
\begin{eqnarray}
h_{\mu\mu}&=&\frac{2Naz}{r^3}, \qquad h_{0\phi}= 2N(1+\cos\theta).
\end{eqnarray}
These are the non-trivial components of the linearized rotating
NUT metric.

\subsection{The rotating NUT without the delta contributions}
If we set ${\Phi^{0y}}_{x} = -{\Phi^{0x}}_{y}=-
{\Phi^{xy}}_{0}={\Phi^{yx}}_{0}= 0$, the difference with the
previous case appears for:
\begin{eqnarray}
\omega_{0i0}&=& -Na\partial_i \partial_z (\frac{1}{r}), \nonumber \\
\omega_{ijk}&=& -Na [\delta_{ik} \partial_j \partial_z
(\frac{1}{r})-\delta_{jk} \partial_i \partial_z
(\frac{1}{r})+\frac{1}{2} \delta_{zi} \partial_k \partial_j
(\frac{1}{r})- \frac{1}{2} \delta_{zj} \partial_k \partial_i
(\frac{1}{r})].
\end{eqnarray}
This means that:
\begin{eqnarray}
R_{00}= -4 \pi N a \delta(x) \delta(y) \delta'(z), \qquad R_{ij}=
-4 \pi N a\delta_{ij} \delta(x) \delta(y) \delta'(z).
\end{eqnarray}
The electric Einstein tensor has now a non-trivial component:
$$G_{00}=-8\pi N a
\delta(x)\delta(y)\delta'(z).$$ The charges for the solution are
thus $K_0=N$ and $L^{0z}=-Na$. This is thus a solution describing
a point magnetic mass $N$ with in addition a ``boost mass" $-Na$
which can be understood as a dipole of electric masses $M$ and
$-M$ separated by a distance $\epsilon$ in the limit where
$\epsilon\rightarrow 0$ and $L_{0z}=Na=M\epsilon$ is kept
constant. Positivity of energy in General Relativity tells us that
this interpretation should be discarded. We present in section 4
the dual version of this calculation.

The interested reader could eventually wonder about different
combinations of the previous considerations. One could for example
try to interpret the rotating NUT solution with only the delta
contributions and no string contribution (or respectively no
$\Phi_{\mu\nu\rho}$ contributions at all). Following our analysis
this only partially matches the proposal of Miller in
\cite{MILLER} to interpret the Kerr-NUT metric as a Schwarzschild
black hole and an infinite source of angular momentum along the
singularity. Our calculations show that it should also be
supplemented with a magnetic angular momentum when delta
contributions are included (respectively with a dipole of electric
masses in the same limit as previously discussed when no
contributions are taken into account). Dual considerations can
also be implemented following the same ideas as presented at the
end of section 4.

\end{document}